\documentclass[11pt,a4paper]{article}
\pdfoutput=1
\usepackage[utf8]{inputenc}
\usepackage{amsmath,amssymb,graphicx,epsfig,color,float,cancel,cite}
\usepackage{caption}
\usepackage{etoolbox}
\usepackage[margin=1.8cm]{geometry}
\usepackage{setspace}
\newcommand{\h}{\bar{h}}
\newcommand{\z}{\bar{z}}
\newcommand{\T}{\mathcal{T}}
\usepackage{comment}
\usepackage{hyperref}
\hypersetup{
  colorlinks   = true, 
  urlcolor     = blue, 
  linkcolor    = blue, 
  citecolor   = red  
}
\newcommand{\nn}{\nonumber}
\newcommand{\f}{\frac}
\newcommand{\be}{\begin{equation}}
\newcommand{\ee}{\end{equation}}
\newcommand{\p}{\partial}

\title{\LARGE Light transformed gluon correlators in CCFT\\ \vspace{1cm}} 
\date{}

\author{Sourish Banerjee, Rudranil Basu\footnote{p20210001@goa.bits-pilani.ac.in, rudranilb@goa.bits-pilani.ac.in} \\ {\it BITS Pilani, KK Birla Goa Campus, Zuarinagar. Goa India - 403726}, \\ Sayali Atul Bhatkar \footnote{sayali.bhatkar@tifr.res.in}\\ {\it Department of Theoretical Physics,
Tata Institute of Fundamental Research}, \\ {\it Homi Bhabha Rd, Mumbai 400005, India}\\ \vspace{2cm}} 
  
\begin{document}

\maketitle

\begin{abstract} 
	\begin{doublespace}
		In the present work, we study celestial correlators of light transformed gluon operators at tree level. We also discuss the transformation of light transformed operators under the action of 4D translations. The two, three and four-point functions arising from MHV amplitudes in terms of light transformed operators satisfy translation invariance constraints, are non-distributional and contain ordinary CFT power law terms. There is a new channel dependent term in the three point function. We show that the  three-point light transformed correlation function is conformally covariant after contributions from all the three channels are added. We also study the OPE limit of the different channels of the three-point function in an attempt to construct a map between the OPE in the Mellin basis and that in the light transformed one. 
	\end{doublespace}
	
\end{abstract}
\newpage
 \tableofcontents

\section{Introduction}
The common wisdom about a quantum theory of gravity, ushered in by the holographic principle is that it should have a non-gravitating field theory dual living at the asymptotic  boundary\cite{Maldacena:1997re}. While this is an established point of view for AdS space-time for the past two-plus decades, the reflections regarding asymptotic flat space-time are relatively new\cite{Bagchi:2019clu,Bagchi:2016bcd, Bagchi:2016geg}. In the case of four-dimensional bulk asymptotically flat space-time, the dual field theory is proposed to be a conformal one, living on the two-dimensional celestial sphere; hence aptly named as celestial conformal field theory (CCFT) \cite{Pasterski:2016qvg,Pasterski:2017kqt,Kapec:2014opa}. The Lorentz group, $SL(2,C)$ of the bulk isometries act as global conformal transformations on the celestial sphere\cite{Stieberger:2018onx, Strominger:2017zoo,Raclariu:2021zjz, Pasterski:2021rjz}. In addition to this, the requirement of the energy-momentum tensor coming from soft particle Ward identities implies at the tree level that the infinite dimensional conformal group generated by the Virasoro algebra should act as the symmetry of theories on the celestial sphere\cite{Banerjee:2019aoy,Banerjee:2019tam,Banerjee:2018fgd,Banerjee:2017jeg}. However, what sets CCFT apart from ordinary two-dimensional CFT is the uncountably infinite number of supertranslation generators, four among which along with the global $SL(2,C)$ correspond to the bulk Poincare isometry of Minkowski space\cite{Bondi:1962px,Sachs:1962zza,Oblak:2015qia}.

The most natural question that one may ask from a holographic principle for asymptotically flat space-time is regarding scattering amplitudes, and the $S$ matrix gives the obvious observable that one may compute. In its present form, this correspondence, also named aptly as Celestial holography is expected to give us a handle to compute bulk scattering $S$ matrix elements in terms of correlators of CCFT primary operators. Mellin transform of lower point tree level bulk scattering amplitudes have been shown to be the same as CCFT correlators. Mellin amplitudes basically involve boost eigenstates, instead of the traditional momentum of plane-wave ones\cite{Pasterski:2016qvg}\cite{Pasterski:2017ylz}. The CCFT correlators are not only constrained by the global conformal symmetry group, $SL(2,C)$ of the celestial sphere but also the translation symmetries of the global Poincare group\cite{Law:2019glh,Stieberger:2018onx}. These additional constraints make CCFT correlators different from ordinary two dimensional CFT, in a sense that the lower  point correlators are distribution valued rather than following a power law. This has been confirmed recently using a candidate dual theory living on asymptotic null infinity as well \cite{Bagchi:2022emh}. However, efforts are being made to find a different basis in the space of celestial operators so as to make the correlators match with those expected in a two-dimensional CFT \cite{Sharma:2021gcz,Crawley:2021ivb,Fan:2021isc,Fan:2021pbp}. Some prominent candidates in this direction are choosing shadow or light transformation, which is tantamount to a different basis for the CCFT operator algebra. These integral transforms are 
particularly important from the perspective of an operator algebra with better behaved structure functions \footnote{More often than not, at the cost of locality of the operators}. However, in certain cases, eg. three-point CCFT correlators vanish and hence their shadow or light transformed versions too. One then argues in favour of going to the split signature in bulk and hence making the celestial theory Lorentzian to get a non-zero three-point correlator. In the Lorentzian signature, it is natural to study light ray operators, which are not necessarily local, albeit transforming as primary under the global conformal group in the context of the celestial sphere. In the space of operators,  local CCFT primaries, shadow primaries and light primaries of course form an over-complete basis\cite{Kravchuk:2018htv, Simmons-Duffin:2012juh, Dolan:2000ut, Ferrara:1972xe, Fan:2021isc,Donnay:2018neh}. Nonetheless, it calls for understanding the operator algebra of these operators.

As a step in this direction, our aim in the present work is to study correlators of light transformed operators. We also discuss the transformation of light transformed operators under the action of 4D translations. As a concrete example, we choose bulk gluon scattering at tree level and focus on the corresponding CCFT correlators. Our checks for two, three and four-point functions arising from MHV amplitudes in terms of light transformed operators satisfy translation invariance constraints and are non-distributional. For a meticulous check we study three point functions for individual scattering channels. The light transformed three point correlator corresponding to the bulk scattering channel of $12\rightleftarrows 3$ is the exact power law term as seen in ordinary CFT's. However the light transformed three point function for the other two channels 
contain a new term which 
break special conformal covariance of these channels individually 
The anomalous terms get cancelled only when the channels are summed up. 
We comment on the origin of this anomaly via a prescription of line integral for the light transform (in Appendix \ref{AppB}). Moreover we use the three-point functions various channels to obtain the leading OPE between the gluon operators 
and light transformed gluon operators. There exists a direct map between the leading singularity in OPE of gluon operators and the leading singularity in the OPE of light transformed gluon operators. But such a direct map does not exist between the respective OPE coefficients. We derive the OPE of light transformed operators in various channels 
and the OPE coefficients turn out to be channel dependent. With the lesson learnt from the analysis of three point functions, we study the four point functions only for the contributions of all the channels summed up.

The organization of the paper is as follows. In section \ref{Sec2}, we infer that the light transformation of primary operators are primaries themselves and light transformation induces a change in basis in the space of celestial operator algebra, at a formal level for Euclidean signature as well. The action of 4D translations on light transformed operators is obtained. We move on to two-point tree level gluon Mellin amplitude in Section 3 and show explicitly that the light transformed version has the standard form of 2D CFT correlator while still satisfying translation symmetry.  For three-point functions, in section \ref{Sec4}, we used the split signature and derived the light transformed three-point function, both for the colour-stripped as well as the full amplitude with colours. The three-point function does not remain invariant under the exchange of the first two operators. This apparent anomaly is solved when we take into account the full amplitude with colour factors. We inserted light transformation for just one operator in the four-point amplitude, in section \ref{Sec5}. For both the three and four-point functions, we noticed that full amplitude with all gluon colour permutations is required to have a particle exchange symmetric correlation function.

During the finalization of the present draft, we noticed the paper \cite{Hu:2022syq}, which has overlap with some of the results presented in our paper.

\section{Light Transformed Celestial Fields} \label{Sec2}
The basic objects in the celestial CFTs are conformal primary fields that transform covariantly under SL(2,C) transformations. Let us start by reviewing the definition of massless celestial primary fields. We start with the on-shell massless scalar field in momentum space which is denoted by $\phi_p(p^\mu)$ and $p^2=0$. Using stereographic projection, null momenta can be conveniently parameterised as :
\begin{equation}\label{MomRD}
 p^\mu=Eq^\mu = E\left( 1+|z|^2,z+\bar{z},z-\bar{z},1-|z|^2\right). 
 \end{equation}
$(z,\z)$ are complex coordinates on the celestial sphere. The field on the celestial sphere is given by the Mellin transform of the on-shell bulk field. 
\be
\widetilde{\phi}_{\Delta}(z,\bar{z}) =  \int_0^\infty d E \ E^{\Delta -1} \phi_p(E,z,\bar{z}). \nn
\ee
The celestial gluon field is defined by the map \cite{Pasterski:2017kqt} :
\begin{equation}\label{cm}
\begin{split}
   O_{a,\mu,h,\bar{h}}(z,\bar{z}) = \int_0^\infty d E \ E^{\Delta -1}A_{a,\mu}(E,z,\bar{z}) . 
\end{split} 
\end{equation}
$A_{a,\mu}$ is a gluon field where we have suppressed the colour index. It satisfies the Lorenz gauge condition, $\p^{\mu}A_{a,\mu}=0$. Conformal dimensions are given by $2h=\Delta+s$, $2\h=\Delta -s$ and $s=\pm 1$. The conformal primaries form a complete basis for $\Delta = 1 + i \lambda$ \cite{Pasterski:2017kqt} i.e. $\Delta$ lies on the principal unitary series. The index $a$ takes values $(+,-)$ and corresponds to two physical helicity states of the gluon.

An n point amplitude can be denoted by $A_{n}(E_j, z_j, \bar z_j)$ where $j$ denotes the momentum of the $j$th scattering particle. The conformal correlator is obtained by Mellin transform of the amplitude :
\begin{equation}
\tilde A_{n}(\Delta_j, z_j, \bar z_j)=\prod_{j=1}^n\int_0^{\infty} dE_j 
E_j^{\Delta_j-1} A_{n}(E_j, z_j, \bar z_j).  \label{3}
\end{equation}
Such correlators have been explicitly obtained for massive or massless operators for various spins \cite{Pasterski:2017ylz, Pasterski:2016qvg,Banerjee:2017jeg,Schreiber:2017jsr} leading to increasing confidence between the celestial correlator - bulk amplitude correspondence.

An undesirable feature of the lower point correlators for massless cases is that they are distribution valued in $(z,\bar{z})$ hence they are ultra-local on the celestial sphere. This led to the study of an alternative set of operators, in the form of shadow or light transform such that the correlators in the new set resemble the usual CFT correlators\cite{Crawley:2021ivb}. Light transformed operators are the main focus of this paper. Although primarily defined for Lorentzian CFTs, we work with a formal definition of light transform of a spin one conformal primary $O_a$ which is defined via the following equation \cite{Sharma:2021gcz,Kravchuk:2018htv,Atanasov:2021cje}
\be 
L^+[{O}]_{a}(w,\bar{w})=\int dz \f{1}{(w-z)^{2-2h}}\ O_{a,h,\bar{h}}(z,\bar{w}).\label{Lp}
\ee
\be 
L^-[{O}]_{a}(w,\bar{w})=\int d\z \f{1}{(\bar{w}-\z)^{2-2\h}}\ O_{a,h,\bar{h}}(w,\bar{z}).\label{Lm}
\ee
In Euclidean signature, we will assume that the above integral over real $z$ can be consistently deformed to a contour encircling the point $w$. The $L^+$-transform is well suited to study positive helicity state while the $L^-$-transform is natural to describe negative helicity state. It should be noted that since $O_a$ is a spin one conformal primary it has $h-\h=\pm 1$ and $\Delta=1+i\lambda$ as it lies on the principal series. Using the contour around $w$, it is a straightforward exercise to check the transformation of the $L^{\pm}$ operators under the SL(2,C) group. The $L^+[{O}]_{a}$-operator transforms as a primary field with $h_L=1-h$ and $\h_L=\h$. Similarly, the $L^-[{O}]_{a}$-operator transforms as a primary field with $h_L=h$ and $\h_L=1-\h$.

Let us consider the following light transformed operator of positive helicity states for concreteness.
\be 
L^+[{O}]_{+}(w,\bar{w})=\int  \f{dz}{(w-z)^{2-2h}}\ O_{+,h,\bar{h}}(z,\bar{w}).\label{Lm}
\ee
The scaling dimension of the above operator is $\Delta_L=0$ and spin is $J=-i\lambda$. Thus these operators have complex-valued continuous spins.  Though the idea of continuous complex spin is unclear at the level of physical understanding, we discuss some desirable features of these operators in this paper. Let us work in Euclidean signature so that the integral is over a contour encircling the point $w$. We consider the cases when $2-2h$ takes positive integral values then the above integral can be evaluated easily using the residue theorem. We get
\begin{align} \label{residue}
L^+[{O}_{h,\h}](w,\bar{w})&=\f{2\pi i}{(1-2h)!}\lim_{z\rightarrow w}\f{\p^{1-2h}}{\p z^{1-2h}}\ O_{h,\bar{h}}(z,\bar{w})\nn.
\end{align}
The above expression holds for $h=\f{1}{2},0,-\f{1}{2},-1,\f{-3}{2},...$ . It is interesting to discuss the light transform of a light transformed operator:
\begin{align}
L^+[L^+[{O}_{h,\h}]](w,\bar{w})&=\oint_w  \f{dz}{(w-z)^{2-2h_L}}\ L^+[O_{h,\bar{h}}](z,\bar{z}) .
\end{align}
Using \eqref{residue}, we arrive at:
%
\begin{align}
L^+[L^+[{O}_{h,\h}]](w,\bar{w})&=\ 2\pi i\oint_w  \f{dz}{(w-z)} \ O_{h,\bar{h}}\ = \ O_{h,\bar{h}}(z,\bar{w}).
\end{align}
Thus we see that the light transform of the light transformed operator is the original operator itself. This shows that light transform is a change of basis in the operator space of the CFT. It remains to be understood if the light transformed operators satisfy state operator correspondence and if the Hilbert space of the CCFT can alternatively be described in terms of states corresponding to the light transformed operators.

An important point to be noted is that the action of four dimensional translations on the light transformed operators is expected to be very different from that of the original operators. Next, we will obtain the transformation of light transformed operators under four dimensional translations. To do so we will assume $2- 2h$ and $2 -2 \h$ to take integral values and then assume that the results can be analytically extended to the case of celestial fields with complex valued conformal dimensions. Under four dimensional translations, the celestial primaries transform in the following way \cite{Stieberger:2018onx}
$$P_\mu O_{a,h,\bar{h}}(z,\bar{z}) = q_\mu\ O_{a,h+\f{1}{2},\bar{h}+\f{1}{2}}(z,\bar{z}).$$
$q_\mu$ has been defined in \eqref{MomRD}. Hence for the light transformed operator, we get
\begin{align}
\mathcal{P}_\mu L^+[{O}_{h,\h}](w,\bar{w})&=\oint_w dz \f{1}{(w-z)^{2-2h}}\ q_\mu\ O_{h+\f{1}{2},\bar{h}+\f{1}{2}}(z,\bar{w}).\label{100}
\end{align}
Again we assume that the above integral over real $z$ can be consistently deformed to be a contour encircling the point $w$. For $2-2h$ taking positive integral values, using the residue theorem for $\mu =0$
\begin{align}
\mathcal{P}_0L^+[{O}_{h,\h}](w,\bar{w})&=\f{2\pi i}{(1-2h)!}\lim_{z\rightarrow w}\ \f{\p^{1-2h}}{\p z^{1-2h}}\ [ (1+z\z)\ O_{h+\f{1}{2},\bar{h}+\f{1}{2}}(z,\bar{z})]\nn
\end{align}
Here we have used $q_0=-(1+z\z)$. Evaluating the derivatives we get
\begin{align}
\mathcal{P}_0L^+[{O}_{h,\h}](w,\bar{w})
&= \f{(1+w\bar{w})}{1-2h}\ \p_w L^+[{O}_{h+\f{1}{2},\bar{h}+\f{1}{2}}](w,\bar{w})+\bar{w}L^+[{O}_{h+\f{1}{2},\bar{h}+\f{1}{2}}](w,\bar{w}).
\end{align}
By similar logic, the action of the momentum operator on light transformed operators can be written as :
\begin{align}
\mathcal{P}_\mu L^+[{O}_{h,\h}](w,\bar{w})
&= \f{q_\mu(w)}{1-2h}\ \p_w L^+[{O}_{h+\f{1}{2},\bar{h}+\f{1}{2}}](w,\bar{w})+[\p_wq_\mu]L^+[{O}_{h+\f{1}{2},\bar{h}+\f{1}{2}}](w,\bar{w}).\label{P+}
\end{align}
\begin{align}
\mathcal{P}_\mu L^-[{O}_{h,\h}](w,\bar{w})
&= \f{q_\mu(w)}{1-2h}\ \p_{\bar{w}} L^-[{O}_{h+\f{1}{2},\bar{h}+\f{1}{2}}](w,\bar{w})+[\p_{\bar{w}}q_\mu]L^-[{O}_{h+\f{1}{2},\bar{h}+\f{1}{2}}](w,\bar{w}). \label{P-}
\end{align}
This is an important result of this paper. This is how the light ray operators transform under the 4D translations. It is non trivial and quite  different from that of the usual celestial operators. We will use this result to check the transformation of light transformed correlators under 4D translations.    
\section{Two Point Function of Gluons } \label{Sec3}
Two-point functions of light transformed gluons have been discussed earlier \cite{Sharma:2021gcz}. We aim to establish the translation invariance of this two-point function. We begin by reviewing the complete derivation. 

To find the two-point function of the gluon primaries we must find the inner product between the corresponding bulk modes and then perform a Mellin transform as in \eqref{3}. One can define a conserved inner product between the on-shell bulk fields as \cite{Donnay:2018neh}, 
\begin{equation}
	(A,A')=\int_{\Sigma} d^{3}X^{i}(A^{j}F'^{*}_{0j}-A'^{j*}F_{0j})
\end{equation}
where we supressed the $a$ index of the fields. Also, $F_{0j}= \p_0 A_{j}-\p_j A_{0} $ and the integration is over a space-like hypersurface $\Sigma$. This inner product is conserved in the sense that it does not depend on the choice of the space-like hypersurface. The on-shell gluon field in momentum space can be written as  $A^{\pm}_{a,\mu}= (\p_a q_\mu)e^{ \pm iEq\cdot X}$ where +/- represent outgoing/incoming waves and $X_{\mu}$ is the position coordinate in 4D Minkowski space.  It was shown in \cite{Pasterski:2017kqt} that the inner product of such single-particle states are,
\begin{equation}
	(\partial_{a_{1}}q_{1}e^{\pm iE_{1}q_{1}\cdot X},\partial_{a_{2}} q_{2}e^{\pm iE_{2}q_{2}\cdot X})= \pm 4 (2\pi)^3E_{1}(1+z_{1}\bar{z_{1}})\delta_{a_{1},a_{2}}\delta^{(3)}(E_{1}\vec{q_{1}}-E_{2}\vec{q_{2}}).
\end{equation}
We consider all the particles to be outgoing particles while some of them are having negative energies i.e we consider only the positive sign in the above equation. On performing the Mellin transform of this amplitude we get the following two-point correlator upto a normalization,
\be
\tilde A_{s_{1},s_{2}} (\Delta_i, z_i, \bar z_i)=  \ \pi \delta_{s_{1},-s_{2}}\delta (\Delta_1+\Delta_2-2)\ \delta (z_{12})\ \delta (\z_{12}).
\ee

Thus the two-point function between a positive helicity gluon primary and a negative helicity gluon primary is given by, 
\begin{align}
\langle{O}_{-}(z_1,\z_1)\ {O}_+(z_2,\z_2) \rangle = \pi \delta (\Delta_1+\Delta_2-2)\ \delta (z_{12})\ \delta (\z_{12}).\label{2pt}
\end{align}
$\Delta_1=1+i\lambda_1,\Delta_2=1+i\lambda_2$ are the respective conformal dimensions of the gluon operators. Thus we see that the two-point function does not display the power-law behaviour seen in usual CFTs. Also, the constraint  $\Delta_1=2-\Delta_2$ is unusual.

Next, we will discuss the two-point function of light transformed gluon operators and show that these resemble the usual CFT correlators. The definition of light transform of spin 1 field has been discussed in \eqref{Lp}. Hence light transforming \eqref{2pt}, we get
\begin{align}
\langle L^-[{O}]_{-}(z_1)\ L^+[{O}]_+(z_2) \rangle\ \ &=\ \oint_{\bar{z}_1}\oint_{z_2} d\z'_1\  dz'_2\ \f{ \ \tilde A_{-+}(z'_1,z'_2)}{(\bar{z_1}-\bar{z}'_1)^{2-2\bar{h}_1}(z_2-z'_2)^{2-2h_2}}. \nn\\
& =\  \f{ \pi \delta (\Delta_1+\Delta_2-2)\ }{(\bar{z_1}-\bar{z}_2)^{2-2\bar{h}_1}(z_2-z_1)^{2-2h_2}}. \label{2L}
\end{align}\\
Here it must be noted that $\delta(z_{12})$ and $\delta(\bar{z}_{12})$ are Delta functions with complex arguments which are well understood as two dimensional Dirac delta functions. However, owing to the (1,3) signature our integration is over a contour as discussed in Section \ref{Sec2}. So, we use a precsription for the Dirac delta function motivated by \cite{Donnay:2020guq} where we express the Dirac delta function in terms of the Gamma function and evaluate the integral. The explicit evaluation of the integral has been in Appendix \ref{AppA1}. Now, as discussed in equation \eqref{Lp} and equation \eqref{Lm}, the conformal dimensions of the light transformed operators are given by \be h_L=\lbrace  \f{i\lambda_1}{2},- \f{i\lambda_2}{2}\rbrace \text{ and } \ \bar{h}_L=-h.\label{hl}\ee
This agrees with \cite{Sharma:2021gcz}. It is important to note that \eqref{2L} is consistent with the expected form of the two-point function of operators with dimensions as given above. It has the required power-law behaviour and it is non zero only for operators with $h_{L1}=h_{L2}$ and $\h_{L1}=\h_{L2}$.

Though it is obvious that \eqref{2L} enjoys conformal covariance, it is not clear if it transforms appropriately under 4D translations and has not been discussed earlier. Denoting translation generators by $\mathcal{P}$ and using \eqref{P-} and \eqref{P+} translation Ward identity implies 
\begin{align}
&[\mathcal{P}_{1\mu}-\mathcal{P}_{2\mu}]  \langle L^-[{O}]_{-}(z_1)\ L^+[{O}]_+(z_2)\  \rangle\nn\\
=&\ \f{q_{1\mu}}{1-2h_1}\p_{\bar{z}_1} \langle L^-[{O}_{h+\f{1}{2},\bar{h}+\f{1}{2}}](z_1,\bar{z}_1)L^+[{O}_{h,\bar{h}}](z_2,\bar{z}_2) \rangle \ -\ \f{q_{2\mu}}{1-2h_2}\p_{z_2} \langle L^-[{O}_{h,\bar{h}}](z_1,\bar{z}_1)L^+[{O}_{h+\f{1}{2},\bar{h}+\f{1}{2}}](z_2,\bar{z}_2) \rangle \ \label{301}
\end{align}
As discussed earlier, translations shift the conformal primaries off the principle series and we need to use the corresponding two-point function,
\begin{align}
\langle L^-[{O}_{h+\f{1}{2},\bar{h}+\f{1}{2}}](z_1,\bar{z}_1)L^+[{O}_{h,\bar{h}}](z_2,\bar{z}_2) \rangle  \ =\f{\pi\delta(\Delta_1+\Delta_2-1)}{(\bar{z_1}-\bar{z}_2)^{1-2\bar{h}_1}(z_2-z_1)^{2-2h_2}}.
\end{align}
It should be recalled that the conformal dimensions of the operators are given by $h_{L^{-}}=({\frac{i\lambda_1}{2},1+\frac{i\lambda_2}{2}}) $ and $\bar{h}_{L^{-}}=(-\frac{i\lambda_1}{2},\frac{i\lambda_2}{2})$. Consequently the above two-point function also displays the correct  power-law behaviour like \eqref{2L}. Again a subtlety here is that the argument of Dirac delta function is a complex quantity hence this function is a generalization of the Dirac delta function discussed in Appendix \ref{AppA1} to be 
\footnote{
$$\phi_{\Delta}=\int _{1-i\infty}^{1+i\infty}(-idz)\delta(i(\Delta-z))\ \phi(z).$$
For Re($\Delta)<1$ we have 
\be
\delta(\Delta-z)=\frac{1}{2\pi}\lim_{\nu\rightarrow 0} \nu^{z-\Delta}\ \Gamma(\Delta-z)
\ee
}
\be
\delta(\Delta_1+\Delta_2-1)=\lim_{\nu\rightarrow 0} \nu^{1-\Delta_1-\Delta_2}\Gamma(\Delta_1+\Delta_2-1)
\ee
Using the above expressions in \eqref{301}, a straightforward algebra leads us to
\begin{align}
&[\mathcal{P}_{1\mu}-\mathcal{P}_{2\mu}] \langle  L^-[{O}]_{-}(z_1)\ L^+[{O}]_+(z_2)\ \rangle  \nn\\
=& \f{\pi \delta(\Delta_1+\Delta_2-1)}{(\bar{z_1}-\bar{z}_2)^{2-2\bar{h}_1}(z_2-z_1)^{2-2h_2}}\Big[-q_{1\mu}+q_{2\mu}+[\p_{\bar{z}_1}q_{1\mu}]\z_{12}-[\p_{z_2}q_{2\mu}]z_{21}\Big]=0.\\\nn
\end{align}
Thus we get :
\be 
[\mathcal{P}_{1\mu}-\mathcal{P}_{2\mu}] \langle L^-[{O}]_{-}(z_1)\ L^+[{O}]_+(z_2)\rangle =0.
\ee
As expected the two-point function is invariant under 4D translations. This is a non-trivial check on our calculations. An interesting point to note about the light transformed operators is that the form of two-point function is fixed by SL(2,C) symmetry and there are no further constraints arising from 4D translation symmetry.

	\section{Three Point Function of Gluons} \label{Sec4}

We need to work in the $(-+-+)$ signature for the three-point amplitude of gluons to be non zero. As given in \cite{Pasterski:2017ylz}, the three-point amplitude of gluons is given by
$$
A_{--+} (E_i, z_i, \bar z_i)= -2\f{E_1E_2}{E_3} \delta^4 (\sum_{i=1}^3\epsilon_i p_i) \f{z_{12}^3}{z_{23}z_{31}}.\label{3bulk}
$$
where $\epsilon_i=1 (-1)$ for outgoing (incoming) particles. 
The conformal correlator is obtained by Mellin transform of the amplitude :
\begin{equation}
	\tilde A_{--+}(\lambda_j, z_j, \bar z_j)=\prod_{j=1}^3\int_0^{\infty} dE_j
	E_j^{i \lambda_j} A_{--+}(E_j, z_j, \bar z_j),
\end{equation}
that turns out to be 
\begin{align}
	\tilde A_{--+}&=-\pi\delta(\sum_j\lambda_j)\ \prod_{j=1}^3
	\sigma_{j*}^{i \lambda_j}\ \frac{\delta(\bar{z}_{12})\delta(\bar{z}_{13})}{\sigma^2_3D^2}\f{z_{12}^3}{z_{23}z_{31}}\ \prod_j\T(\sigma_{*j})\label{26}
\end{align}
Here we have used the notation
\begin{align*}
	&\sigma_1^* =  \frac{z_{23}}{D }, \sigma_2^* = -  \epsilon_1\epsilon_2\frac{z_{13}}{D },
	\sigma_3^* =  \epsilon_1\epsilon_3 \frac{z_{12}}{D },\nn\\
	& D = (1-\epsilon_1\epsilon_2) z_{13}+ (\epsilon_1\epsilon_3-1)z_{12},
\end{align*}
and $\T(x)$ is the indicator function defined by :
\begin{align} \T(x) &:= 1 \ \ x\ \epsilon\ [0,1]\nn\\
	& :=0 \ \text{ otherwise}.\end{align}
The conformal dimensions of gluon operators are given by $$ h_1= \frac{i\lambda_1}{2}, h_2= \frac{i\lambda_2}{2}, h_3= 1+\frac{i\lambda_3}{2}$$
$$ \bar{h}_1= 1+\frac{i\lambda_1}{2}, \bar{h}_2= 1+\frac{i\lambda_2}{2}, \bar{h}_3= \frac{i\lambda_3}{2}.$$
\subsection{Light Transformed Three Point Function}

Next, we will use equation \eqref{26} to obtain the three-point functions of light transformed operators. Since the first and the second operators have negative helicities, let us use the second kind of light transform \eqref {Lm} for the first two operators. We will use the first kind of light transform \eqref{Lp} for the third operator. We get,
\begin{align}\langle L^-[{O}]_{-}(z_1)\ L^-[{O}]_-(z_2)\ L^+[{O}]_+(w)\rangle\ \ &=\ \int d\z'_1\ d\z'_2\ dz'_3\ \f{ \ \tilde A_{--+}(z'_1,z'_2,z'_3)}{(\bar{z_1}-\bar{z}'_1)^{2-2\bar{h}_1}}\f{1}{(\bar{z_2}-\bar{z}'_2)^{2-2\bar{h}_2}(z_3-z'_3)^{2-2h_3}}. \label{101}
\end{align}

\subsubsection{ The $12\rightleftarrows 3$ channel}

For concreteness we first consider $12\rightarrow 3$ channel i.e. $\epsilon_1=\epsilon_2=-\epsilon_3=-1$. Using \eqref{26} for $12\rightleftarrows 3$ : 
\begin{align}
	\tilde A_{--+}|_{12\rightarrow 3}
	&=-\pi\delta(\sum_j\lambda_j)\ z_{23}^{i \lambda_1} z_{13}^{i \lambda_2}z_{12}^{i \lambda_3}\ \delta(\bar{z}_{12})\delta(\bar{z}_{13})\f{z_{12}}{z_{23}z_{13}}\ \prod_j\T(\sigma_{*j})|_{12\rightarrow 3}.\label{100}
\end{align} 
Next we use the 'indicator' functions' $\mathcal{T}$ we get following conditions :\\
\hspace*{1cm} For the case $z_{12}>0$ : $z_1 < z_3' < z_2$ and\\ 
\hspace*{1cm} for the case $z_{12} < 0$ : $z_2 < z_3' < z_1$.\\
Using the 'indicator' functions' $\mathcal{T}$ in \eqref{100}, the $z_3'$ integral gives us
\begin{align}
	\int_{z_2}^{z_1} dz'_3 \f{sgn(z_{12})}{(z_3-z_3')^{-i \lambda_3}}\ z_{23'}^{-1+i \lambda_1} z_{13'}^{-1+i \lambda_2}z_{12}^{1+i \lambda_3}\  = -{z_{23}^{i \lambda_3}}\  \Gamma(i\lambda_1)\ \Gamma(i\lambda_2) \ {}_2{F}_1[i\lambda_1,-i\lambda_3,i\lambda_1+i\lambda_2,\f{z_{12}}{z_{32}}]. 
\end{align}\label{123}
Here ${}_2F_1$ is the usual hypergeometric function. The integrals over the delta functions can be done trivially to arrive at 
\begin{align}
	&\langle L^-[{O}]_{-}(z_1)\ L^-[{O}]_-(z_2)\ L^+[{O}]_+(z_3)\rangle\ |_{12\rightarrow 3}\ \nn\\
	&=\pi\ sgn(z_{12})\ \delta(\sum_j\lambda_j)\  {z_{23}^{i \lambda_3}}\ \z_{13}^{i \lambda_1}\ \z_{23}^{i \lambda_2} \  \Gamma(i\lambda_1)\ \Gamma(i\lambda_2) \ {}_2{F}_{1}[i\lambda_1,-i\lambda_3,i\lambda_1+i\lambda_2,\f{z_{12}}{z_{32}}]. 
\end{align}
Next we use the delta function constraint, $\delta(\sum_j\lambda_j)$, to get  $i\lambda_1+i\lambda_2 = -i\lambda_3$ in the argument of the hypergeometric function to get
\begin{align}
	&\langle L^-[{O}]_{-}(z_1)\ L^-[{O}]_-(z_2)\ L^+[{O}]_+(z_3)\rangle\ |_{12\rightarrow 3}\nn\\
	& =-\pi\ sgn(z_{12})\ \delta(\sum_j\lambda_j)\  {z_{23}^{i \lambda_3+i\lambda_1}}\  \z_{13}^{i \lambda_1}\ \z_{23}^{i \lambda_2}\ z_{13}^{-i \lambda_1}\  \f{\Gamma(i\lambda_1)\ \Gamma(i\lambda_2)}{\Gamma(-i\lambda_3)}. \label{3L}
\end{align}
This agrees with the result of \cite{Sharma:2021gcz}.
We note that the $z_{ij}$ factors in the above expression are consistent with the expected form of the three-point function\footnote{
	$$ z_{12}^{-h_1-h_2+h_3}\ z_{23}^{-h_3-h_2+h_1}\ z_{13}^{-h_1-h_3+h_1}\ \times \ c.c.\ .$$}
when we use the fact that the dimensions of the light transformed operators are given by \be h_L=\lbrace  \f{i\lambda_1}{2}, \f{i\lambda_2}{2},- \f{i\lambda_3}{2}\rbrace \text{ and } \ \bar{h}_L=-h_L.\label{hl3}\ee
The absence of $z_{12}$ factor seems strange but it is not so since $z_{12}$ is expected to appear as  $z_{12}^{\f{-i}{2}(\lambda_1+\lambda_2+\lambda_3)}$ which is 1 due to the delta function constraint.
Let us check the translation invariance of the three-point function. We assume equations \eqref{P+} and \eqref{P-} hold in the split signature as well. Consequently we find that the Ward identity for the three-point function to be,
\begin{align}
	&\big(\mathcal{P}_{1\mu}+\mathcal{P}_{2\mu}-\mathcal{P}_{3\mu}\big) \langle L^-[{O}]_{-}(z_1)\ L^-[{O}]_-(z_2)\ L^+[{O}]_+(z_3)\rangle =\nn \\ 
	&z_{32}^{-2h_2 -1}  \z_{31}^{2h_1-1}\ \z_{23}^{2\h_2-2}\ \z_{13}^{2\h_1-2}\  \f{\Gamma(2h_1)\ \Gamma(2h_2)}{\Gamma(2-2h_3)}\nn\\
	&  \times \Big[2h_{1}z_{32}\big(-q_{3\mu}+q_{1\mu}-[\p_{\bar{z}_1}q_{1\mu}]\z_{13}+[\p_{z_3}q_{3\mu}]z_{31}\big) \nn \\&+2h_{2}z_{31}\big(-q_{3\mu}+q_{2\mu}-[\p_{\bar{z}_2}q_{2\mu}]\z_{23}+[\p_{z_3}q_{3\mu}]z_{32}\big)\big)\Big]=0.
\end{align}
This follows from the result of section \ref{Sec3}. We also note that the \eqref{3L} satisfies the $SL(2, \mathbb{R}) \times SL(2, \mathbb{R})$ global conformal Ward identities upto a $sgn$ function. 

An interesting point to note is the presence of sgn($z_{12}$) in \eqref{3L}. Such discontinuous behaviour is not seen in CFT correlators. CFT correlators are expected to be symmetric under the exchange of $1\leftrightarrow 2$, but the presence of sgn($z_{12}$) in \eqref{3L} makes the above three-point function antisymmetric under this exchange. This property is inherited by the three-point function from the bulk amplitude in \eqref{3bulk}. To construct crossing symmetric correlators we have to take into account the colour factors  as well. This will be done in subsection \ref{Sec 4.3}.

\subsubsection{ The $13\rightleftarrows 2$ channel}

Let us consider a different channel such that $13\rightarrow 2$. The Mellin amplitude for this channel is given by 
\begin{align}
	\tilde A_{--+}|_{13\rightarrow 2}
	&=\pi\delta(\sum_j\lambda_j)\ \ z_{23}^{-1+i \lambda_1} z_{13}^{-1+i \lambda_2}z_{12}^{1+i \lambda_3}\ \delta(\bar{z}_{12})\delta(\bar{z}_{13})\ \prod_j\T(\sigma_{*j})|_{13\rightarrow 2}.\nn
\end{align} 
The 'indicator' functions' $\mathcal{T}$ in this channel lead to following conditions :\\
\hspace*{1cm} For the case $z_{12}<0$ : $z_1  < z_2< z_3'$ and\\ 
\hspace*{1cm} for the case $z_{12} > 0$ : $z_3' <z_2 <  z_1$.\\
For the case $z_{12}<0$, the $z_3'$ integral gives us
\begin{align}
	& \int_{z_2}^{\infty} dz'_3 \f{1}{(z_3-z_3')^{-i \lambda_3}}\ z_{23'}^{-1+i \lambda_1} z_{13'}^{-1+i \lambda_2}\nn\\
	&  = - z_{13}^{-i \lambda_1}\ {z_{23}^{-i \lambda_2}}\ z_{12}^{-1+i \lambda_2+i\lambda_1} \f{ \Gamma(i\lambda_1)\ \Gamma(i\lambda_2)}{\Gamma(i\lambda_1+i\lambda_2)} \ -\ \f{1}{z_{12}}  \f{  \Gamma(-i\lambda_2)}{\Gamma(1-i\lambda_2)}{}_2{F}_1[1,i(\lambda_1+\lambda_2),1+i\lambda_2,\f{z_{32}}{z_{12}}]. \label{132}
\end{align}
Similar calculation can be done for $z_{12}>0$. Generalizing for both $z_{12}>0$ and $z_{!2}<0$ we find that the light transformed correlator is, 
\begin{align}
	&\langle L^-[{O}]_{-}(z_1)\ L^-[{O}]_-(z_2)\ L^+[{O}]_+(z_3,\z_3)\rangle\ |_{13\rightleftarrows 2}\ \nn\\
	&=\pi\ sgn(z_{12}) \delta(\sum_j\lambda_j)\   \z_{13}^{i \lambda_1}\ \z_{23}^{i \lambda_2} \Big[ z_{13}^{-i \lambda_1} {z_{23}^{-i \lambda_2}}  \f{ \Gamma(i\lambda_1)\ \Gamma(i\lambda_2)}{\Gamma(i\lambda_1+i\lambda_2)} + z_{12}^{i \lambda_3}  \f{  \Gamma(-i\lambda_2)}{\Gamma(1-i\lambda_2)}{}_2{F}_1[1,i(\lambda_1+\lambda_2),1+i\lambda_2,\f{z_{32}}{z_{12}}]\ \Big]\label{L132}
\end{align}
The first term in above expression is the exact analogue of the $12\leftrightarrows 3$ correlator in \eqref{3L}. Hence we expect that it satisfies all the Ward identities by itself as we checked for the 
$12\leftrightarrows 3$ case. The appearance of the second term is quite surprising. Thus there is a new three-point structure in celestial correlators. 

Let us explicitly check the transformation of the new term :
\be f(h_1,h_2,h_3) =  \z_{13}^{i \lambda_1}\ \z_{23}^{i \lambda_2} z_{12}^{i \lambda_3}  \f{  \Gamma(-i\lambda_2)}{\Gamma(1-i\lambda_2)}{}_2{F}_1[1,i(\lambda_1+\lambda_2),1+i\lambda_2,\f{z_{32}}{z_{12}}], \label{f} \ee
where
\be {}_2{F}_1[1,i(\lambda_1+\lambda_2),1+i\lambda_2,\f{z_{32}}{z_{12}}] = \left( \f{z_{13}}{z_{12}}\right)^{-i\lambda_1} {}_2{F}_1[i\lambda_2,1-i\lambda_1,1+i\lambda_2,\f{z_{32}}{z_{12}}]. \ee
We find that although it satisfies the Ward identities for translation and dilatation, it fails to do so for the Ward identity corresponding to special conformal transformation. The $f$ term is not eliminated by the action of SCT generators :
\be \sum_{i=1}^3[z^2_i\p_{z_i}+2h_iz_i]\ f(h_1,h_2,h_3)=-i\lambda_2 \f{  \Gamma(-i\lambda_2)}{\Gamma(1-i\lambda_2)}(z_1-z_2)^{i\lambda_3+1}=(z_1-z_2)^{i\lambda_3+1}.\label{sct}\ee
This makes the light transformed three-point function representing the $13\rightleftarrows 2$ channel of gluon amplitude non-covariant under the global conformal group. This point has been elaborated in Appendix \ref{AppB}. The explicit calculation for \eqref{sct} has been done in Appendix \ref{AppC}.

\subsubsection{ The $23\rightleftarrows 1$ channel}

he Mellin amplitude for this channel is given by 
\begin{align}
	\tilde A_{--+}|_{23\rightarrow 1}
	&=-\pi\delta(\sum_j\lambda_j)\ \ z_{23}^{-1+i \lambda_1} z_{13}^{-1+i \lambda_2}z_{12}^{1+i \lambda_3}\ \delta(\bar{z}_{12})\delta(\bar{z}_{13})\ \prod_j\T(\sigma_{*j})|_{23\rightarrow 1}.\nn
\end{align} 
The 'indicator' functions' $\mathcal{T}$ in this channel lead to following conditions :\\
\hspace*{1cm} For the case $z_{12}<0$ : $z_3'<z_1  < z_2 $ and\\ 
\hspace*{1cm} for the case $z_{12} > 0$ : $z_2 <  z_1<z_3'$.\\
For the case $z_{12}<0$, the $z_3'$ integral gives us
\begin{align}
	& \int^{z_1}_{-\infty} dz'_3 \f{1}{(z_3-z_3')^{-i \lambda_3}}\ z_{23'}^{-1+i \lambda_1} z_{13'}^{-1+i \lambda_2}\nn\\
	&  = (-1)^{-i\lambda_1}\ z_{23}^{-i \lambda_2}\ {z_{13}^{-i \lambda_1}}\ z_{21}^{-1+i \lambda_2+i\lambda_1} \f{ \Gamma(i\lambda_1)\ \Gamma(i\lambda_2)}{\Gamma(i\lambda_1+i\lambda_2)} \ -\ \f{1}{z_{12}}  \f{  \Gamma(-i\lambda_1)}{\Gamma(1-i\lambda_1)}{}_2{F}_1[1,i(\lambda_1+\lambda_2),1+i\lambda_1,\f{z_{13}}{z_{12}}]. \label{231}
\end{align}
Like the previous section we have the following general expression for the light transform of the correlator, 
\begin{align}
	&\langle L^-[{O}]_{-}(z_1)\ L^-[{O}]_-(z_2)\ L^+[{O}]_+(z_3,\z_3)\rangle\ |_{23\rightarrow 1}\ \nn\\
	&=\pi\ sgn(z_{12}) \delta(\sum_j\lambda_j)\   \z_{13}^{i \lambda_1}\ \z_{23}^{i \lambda_2} \Big[z_{23}^{-i \lambda_2}\ {z_{13}^{-i \lambda_1}}\  \f{ \Gamma(i\lambda_1)\ \Gamma(i\lambda_2)}{\Gamma(i\lambda_1+i\lambda_2)} + (z_{12})^{i\lambda_3}  \f{  \Gamma(-i\lambda_1)}{\Gamma(1-i\lambda_1)}{}_2{F}_1[1,i(\lambda_1+\lambda_2),1+i\lambda_1,\f{z_{13}}{z_{12}}]\ \Big]\label{3L231}
\end{align}
We again have a new structure in the light transformed three-point function, 
\be f_2= \ (z_{12})^{i\lambda_3}  \f{  \Gamma(-i\lambda_1)}{\Gamma(1-i\lambda_1)}{}_2{F}_1[1,i(\lambda_1+\lambda_2),1+i\lambda_1,\f{z_{13}}{z_{12}}] \label{f2} \ee
For the generator of special conformal transformations  
\be \sum_{i=1}^3[z^2_i\p_{z_i}+2h_iz_i]\ f_2(h_1,h_2,h_3)=i\lambda_1\f{  \Gamma(-i\lambda_1)}{\Gamma(1-i\lambda_1)}(z_1-z_2)^{i\lambda_3+1}=-(z_1-z_2)^{i\lambda_3+1}. \label{sct1}
\ee
\color{black}
We see that \eqref{sct} cancels \footnote{in \eqref{sct1}, we have chosen a phase convention, that is $(-1)^{-i\lambda}=(-1)^{i\lambda}$, as in \cite{Pasterski:2017ylz} and \cite{Fan:2021isc}.} exactly with \eqref{sct1}. Since the the $12\rightleftarrows 3$ channel on its own is conformally covariant, we conclude that the light transformed three point function representing scattering of 3 gluons is conformally covariant only if all the channels are added up.

\subsection{OPE}
In this section we discuss the OPE of two Mellin transformed operators by taking the OPE limit of the various channels of the three-point function. We also write the OPE for the Light transformed operators and find their dependece on the various channels.

\subsubsection{12$\rightleftarrows$3 channel}

For this channel we have the following expression for the Mellin transformed amplitude, 
\begin{align}
	\langle O^{a}_{\Delta_1,-}(z_1) O^{b}_{\Delta_2,-}(z_2) O^{c}_{\Delta_3,+}(z_3)\rangle|_{12\rightleftarrows 3} =\pi f^{abc} \delta(\sum_{j=1}^{3}\lambda_{j}) z_{23}^{i\lambda_1-1}z_{13}^{i\lambda_2-1}z_{12}^{1+i\lambda_3}\ \delta(\z_{12})\ \delta(\z_{23})\ \T(\sigma_{*j})|_{12\leftrightarrows 3}\ .\label{n3}
\end{align}
The $\T$ functions can be written as products and sum of the Heaviside step-fuction,
$$ \T|_{12\rightleftarrows 3}= \Theta(z_{13})\Theta(z_{32})\Theta(z_{12})+\Theta(z_{23})\Theta(z_{31})\Theta(z_{21}).$$
We aim to discuss the OPE expansion of above correlator. It has a delta function singularity which cannot be reproduced by finite number of power law singularities. This tells us that there should be a $\delta$ function in the following OPE 
\begin{align}
	O^{a}_{\Delta_1,-}(z_1) O^{b}_{\Delta_2,-}(z_2) = f^{ab}_{c} \  \delta(\z_{12})\ O^{c}_{(\Delta_1+ \Delta_2-1),-}(z_2)+ ...\ .\label{ope}
\end{align}
The conformal  dimension of the operator on the right is fixed using symmetry. Let us discuss the implication of \eqref{ope} for the three-point function of $O^{a}_-(z_1)O^{b}_-(z_2)O^{c}_+(z_3)$. As $1\rightarrow 2$, we expect the following
\begin{align}
	\langle O^{a}_{\Delta_1,-}(z_1) O^{b}_{\Delta_2,-}(z_2) O^{c}_{\Delta_3,+}(z_3)\rangle|_{12\rightleftarrows 3} \ & \rightarrow\ f^{abd} \ sgn(z_{12})\ \delta(\z_{12})\ \langle O^d_{\Delta_1+\Delta_2-1,-}(z_2) O^c_{\Delta_3,+}(z_3)\rangle\ + ...\ .\nn\\
	\ & \rightarrow\ \pi f^{abc} \delta(\lambda_1+\lambda_2+\lambda_3)\ sgn(z_{12})\ \delta(\z_{12})\ \delta^2(z_{23})\ +...\ .
\end{align}
But such delta functions are not seen in \eqref{n3} which seems contradictory. But we show that this is not the case. Let us study the OPE decomposition of \eqref{n3} in the limit $1\rightarrow 2$ 
\begin{align}
\langle O^{a}_{\Delta_1,-}(z_1) O^{b}_{\Delta_2,-}(z_2) O^{c}_{\Delta_3,+}(z_3)\rangle|_{12\rightleftarrows 3}  \ \rightarrow\ & \pi f^{abc} \delta(\lambda_1+\lambda_2+\lambda_3) z_{13}^{i\lambda_1+i\lambda_2-2}z_{12}^{1+i\lambda_3}\ \times\nn\\ 
	& \delta(\z_{12})\ \delta(\z_{23})\ [\Theta(z_{13})\Theta(z_{32})\Theta(z_{12})+\Theta(z_{23})\Theta(z_{31})\Theta(z_{21})]
\end{align}
The Theta functions tell us that taking $1\rightarrow 2$ forces all points to come together.
Let us consider the case $z_1>z_3>z_2$, that is $z_{12}>0$, and use : 
\be \Theta(z_{13})  \ \rightarrow\ \Theta(z_{23})\ +\ z_{12}\ \delta(z_{23})\ +\ ...\ . \ee
We get
\begin{align}
\langle O^{a}_{\Delta_1,-}(z_1) O^{b}_{\Delta_2,-}(z_2) O^{c}_{\Delta_3,+}(z_3)\rangle  \rightarrow\ i\pi f^{abc} \delta(\lambda_1+\lambda_2+\lambda_3) z_{13}^{i\lambda_1+i\lambda_2-2}z_{12}^{2+i\lambda_3}\ \delta(\z_{12})\ \delta(\z_{23})\ \Theta(z_{12})\ \delta(z_{23})+...
\end{align}
where we have used the convention $\Theta(0)=1$.
Considering both $z_{12}>0$ and $z_{12}<0$ we get, 
\begin{align}
\langle O^{a}_{\Delta_1,-}(z_1) O^{b}_{\Delta_2,-}(z_2) O^{c}_{\Delta_3,+}(z_3)\rangle &  \rightarrow \pi f^{abc} \delta(\lambda_1+\lambda_2+\lambda_3)\ z_{13}^{i\lambda_1+i\lambda_2-2}z_{12}^{2+i\lambda_3}\ \delta(\z_{12})\ \delta(\z_{23})\ sgn(z_{12})\ \delta(z_{23})+...\nn\\
	& \rightarrow f^{abd}  \delta(\z_{12})\ sgn(z_{12})\ \langle O^{d}_-(z_2) O^{c}_+(z_3)\rangle+...
\end{align}
This is a very non-trivial check on the delta function in \eqref{ope}. Above calculation tells us that \eqref{ope} should be modified to include the $sgn$ function :  
\begin{align}
	O_{\Delta_1,-}^a(z_1)\ O_{\Delta_2,-}^b(z_2) =  f^{ab}_c sgn(z_{12}) \delta(\z_{12})\ O^c_{(\Delta_1+\Delta_2-1),-}(z_2)\label{ope2}
\end{align}
It should be noted that $ sgn(z_{12})$ makes sure that the RHS is invariant under simultaneous exchange of 1$\leftrightarrow$2 and $a\leftrightarrow b$ like the LHS.

Before moving onto the OPE of light transformed operators some comments are in order. The implications of \eqref{ope2} on the higher point functions need to be studied. The power law singularity in above OPE has already been discussed and takes the following form \cite{Pate:2019lpp} :
\begin{align}
	O_{\Delta_1,-}^a(z_1)\ O_{\Delta_2,-}^b(z_2) \ = \  f^{ab}_{c} \ sgn(z_{12}) &  \delta(\z_{12}) O^c_{(\Delta_1+\Delta_2-1),-}(z_2)\nn\\
	&+\f{-if^{ab}_{\ c}}{\z_{12}}\f{\Gamma(\Delta_1-1)\Gamma(\Delta_2-1)}{\Gamma(\Delta_1+\Delta_2-2)}\  {O}^c_{(\Delta_1+\Delta_2-1),-}(z_2)+\ ...\ .\label{mm2}
\end{align}
One question remains unclear : How does the second operator in above expansion contribute to the three point function \eqref{n3}?\footnote{ The expected contribution would be 
	\begin{align}
		\langle O^{a}_-(z_1) O^{b}_-(z_2) O^{c}_+(z_3)\rangle & = -i f^{ab}_{c}\ \f{\Gamma(i\lambda_1)\Gamma(i\lambda_2)}{\Gamma(i\lambda_1+i\lambda_2)} \f{1}{\z_{12}}\ \langle O_- O_+\rangle\nn\\
		&= -if^{ab}_{c} \f{\Gamma(i\lambda_1)\Gamma(i\lambda_2)}{\Gamma(i\lambda_1+i\lambda_2)}\ \f{1}{\z_{12}} \delta(i\lambda_1+i\lambda_2+i\lambda_3) \delta^2(z_{23})\ +...\ .
	\end{align}
This term is not visible in equation \eqref{n3} and we leave this issue for future investigations.} 

Next we study the light transform of the OPE in \eqref{mm2}. It should be made clear at the onset that this is not a consistent operation as the OPE expansion would not be valid for the entire domain of light transform. Nevertheless it is tempting to go ahead. We take the light transform of the delta function singularity in \eqref{ope2} to get
\begin{align}
	L^-[O^a_{\Delta_1,-}](z_1)\ L^-[O^b_{\Delta_2,-}](z_2) &= f^{ab}_c  \ sgn(z_{12})\ f^{ab}_c\  \int \f{d\z}{(\z-\z_2)^{4-2\h_1-2\h_2}}O^c_-(z_2,\z_2)\ +\ ...\nn\\
	&= f^{ab}_c  \ sgn(z_{12})  L^-[O^c_{(\Delta_1+\Delta_2-1),-}](z_2,\bar{z_2})\ +\ ...\ .\label{Lope}
\end{align}
Let us check if this is consistent with the three-point function. We study the $1\rightarrow 2$ limit of \eqref{3L}. 
\begin{align}
	\langle L^-[{O}]_{\Delta_1,-}(z_1)\ L^-[{O}]_{\Delta_2,-}(z_2)\ L^+[{O}]_{\Delta_3,+}&(z_3)\rangle  |_{12\rightarrow 3 } \rightarrow \pi\ sgn(z_{12}) \delta(\sum_{j=1}^{3}\lambda_{j})  {z_{32}^{i \lambda_3}}   \z_{23}^{i \lambda_2}\ \f{\Gamma(i\lambda_1)\ \Gamma(i\lambda_2)}{\Gamma(i\lambda_1+i\lambda_2)}+ ... \nn\\
	&\rightarrow \ sgn(z_{12})\ \f{\Gamma(i\lambda_1) \Gamma(i\lambda_2)}{\Gamma(i\lambda_1+i\lambda_2)} \langle L^-[{O}]_{(\Delta_1+\Delta_2-1),-}(z_2)\ L^+[{O}]_{\Delta_3,+}(z_3)\rangle +...
\end{align}
Using above expression the OPE turns out to be :
\be
L^-[{O}]_{\Delta_1,-}(z_1)\ L^-[{O}]_{\Delta_2,-}(z_2) =\ sgn(z_{12})\ \f{\Gamma(i\lambda_1)\ \Gamma(i\lambda_2)}{\Gamma(i\lambda_1+i\lambda_2)}\ L^-[{O}]_{(\Delta_1+\Delta_2-1),-}(z_2)\ .\label{lope2}
\ee
This agrees with the singularity structure of \eqref{Lope} but the OPE coefficients do not match. Thus there seems to be a direct map between the leading singularity in OPE of gluon operators and the leading singularity in the OPE of light transformed gluon operators. But such a direct map  does not exist between the respective OPE coefficients.

\subsubsection{13$\leftrightarrows$2 channel}

We follow the analysis of the previous section and take the OPE limit of the Mellin transformed amplitude, 
\begin{align}
	\langle O^{a}_-(z_1) O^{b}_-(z_2) O^{c}_+(z_3)\rangle|_{13\rightleftarrows 2} =\pi f^{abc} \delta(\sum_{j=1}^{3}\lambda_{j}) z_{23}^{i\lambda_1-1}z_{13}^{i\lambda_2-1}z_{12}^{1+i\lambda_3}\ \delta(\z_{12})\ \delta(\z_{23})\ \T(\sigma_{*j})|_{13\leftrightarrows 2}\ .\label{13to2}
\end{align}

On taking the OPE limit we find that upto leading order in $z_{12}$ the correlator has a similar structure as the correlator of the $12\leftrightarrows3$ channel, 
\begin{align}
	\langle O^{a}_{\Delta_1,-}(z_1) O^{b}_{\Delta_2,-}(z_2) O^{c}_{\Delta_3,+}(z_3)\rangle|_{13\rightleftarrows 2} \rightarrow f^{abd}  \delta(\z_{12})\ sgn(z_{12})\ \langle O^{d}_{(\Delta_1+\Delta_2-1),-}(z_2) O^{c}_{\Delta_3,+}(z_3)\rangle+...
\end{align}
which results in the OPE of these two opertaors to have the form,
 \begin{equation}
 		O_{\Delta_1,-}^a(z_1)\ O_{\Delta_2,-}^b(z_2) =  f^{ab}_c sgn(z_{12}) \delta(\z_{12})\ O^c_{(\Delta_1+\Delta_2-1),-}(z_2) \label{OPE13to2}.
 \end{equation}
This is same as \eqref{ope2}. Next we look for the OPE limit of the Light transformed correlator of the $13\leftrightarrows2$ channel. We take $1\rightarrow2$ in equation \eqref{L132} and find that result can be 
recast in terms of the two-point function of the two operators as follows,
\begin{align}
	\langle L^-[{O}]_{\Delta_1,-}(z_1)\ L^-[{O}]_{\Delta_2,-}(z_2)\ L^+[{O}]_{\Delta_3,+}(z_3)\rangle \ |_{13\leftrightarrows 2 } \rightarrow 	sgn(z_{12})\  &\bigg[\f{\Gamma(i\lambda_1)\ \Gamma(i\lambda_2)}{\Gamma(i\lambda_1+i\lambda_2)}\ -  \frac{\Gamma(i\lambda_2)\Gamma(1-i\lambda_1-i\lambda_2)}{\Gamma(1-i\lambda_1)} \bigg] \nn \\  
&	\times \langle L^-[{O}]_{(\Delta_1+\Delta_2-1),-}(z_2)\ L^+[{O}]_{\Delta_3,+}(z_3)\rangle +...
\end{align}
So the OPE takes the form, 
\begin{align}
	L^-[{O}]_{\Delta_1,-}(z_1)\ L^-[{O}]_{\Delta_2,-}(z_2)|_{13\leftrightarrows2} =\ sgn(z_{12})\ \bigg[\f{\Gamma(i\lambda_1)\ \Gamma(i\lambda_2)}{\Gamma(i\lambda_1+i\lambda_2)}\ - \frac{\Gamma(i\lambda_2)\Gamma(1-i\lambda_1-i\lambda_2)}{\Gamma(1-i\lambda_1)} \bigg] L^-[{O}]_{(\Delta_1+\Delta_2-1),-}(z_2)\label{Lope13to2}
\end{align}
\subsubsection{23$\leftrightarrows $1 channel} 
We have the $23\leftrightarrows1$ channel, 
\begin{align}
	\langle O^{a}_{\Delta_1,-}(z_1) O^{b}_{\Delta_2,-}(z_2) O^{c}_{\Delta_3,+}(z_3)\rangle|_{23\rightleftarrows 1} =\pi f^{abc} \delta(\sum_{j=1}^{3}\lambda_{j}) z_{23}^{i\lambda_1-1}z_{13}^{i\lambda_2-1}z_{12}^{1-i\lambda_1-i\lambda_2}\ \delta(\z_{12})\ \delta(\z_{23})\ \T(\sigma_{*j})|_{23\leftrightarrows 1}\ .\label{23to1}
\end{align}


We find that upto leading order in $z_{12}$ the OPE limit of the correlator is, , 
\begin{align}
\langle O^{a}_{\Delta_1,-}(z_1) O^{b}_{\Delta_2,-}(z_2) O^{c}_{\Delta_3,+}(z_3)\rangle|_{23\leftrightarrows 1} \rightarrow f^{abd}  \delta(\z_{12})\ sgn(z_{12})\ \langle O^{d}_{(\Delta_1+\Delta_2-1),-}(z_2) O^{c}_{\Delta_3,+}(z_3)\rangle+...
\end{align}
which reproduces the OPE, 
\begin{equation}
	O_{\Delta_1,-}^a(z_1)\ O_{\Delta_2,-}^b(z_2) =  f^{ab}_c sgn(z_{12}) \delta(\z_{12})\ O^c_{(\Delta_1+\Delta_2-1),-}(z_2) \label{OPE23to1}.
\end{equation}
The OPE limit of the Light transformed correlator of the $23\leftrightarrows1$, 
like the previous section, can be rewritten in terms of the two-point function of the two operators as ,
\begin{align}
\langle L^-[{O}]_{\Delta_1,-}(z_1) L^-[{O}]_{\Delta_2,-}(z_2) L^+[{O}]_{\Delta_3,+}(z_3)\rangle  |_{23\leftrightarrows 1 } \rightarrow 	sgn(z_{12})  \bigg[&\f{\Gamma(i\lambda_1)\ \Gamma(i\lambda_2)}{\Gamma(i\lambda_1+i\lambda_2)}\ -  \frac{\Gamma(i\lambda_1)\Gamma(1-i\lambda_1-i\lambda_2)}{\Gamma(1-i\lambda_2)} \bigg] \nn \\  
	&	\times \langle L^-[{O}]_{(\Delta_1+\Delta_2-1)}(z_2)\ L^+[{O}]_{\Delta_3,+}(z_3)\rangle +...
\end{align}
Hence,the OPE takes the form, 
\begin{align}
	L^-[{O}]_-(z_1)\ L^-[{O}]_-(z_2) |_{23\leftrightarrows1} = sgn(z_{12}) \bigg[\f{\Gamma(i\lambda_1) \Gamma(i\lambda_2)}{\Gamma(i\lambda_1+i\lambda_2)} - \frac{\Gamma(i\lambda_1)\Gamma(1-i\lambda_1-i\lambda_2)}{\Gamma(1-i\lambda_2)} \bigg] L^-[{O}]_{(\Delta_1+\Delta_2-1),-}(z_2)\label{Lope23to1}
\end{align}

In this subsection we obtained the leading term in OPE of celestial gluon operators using the respective three point functions. Using the $12 \leftrightarrows 3$ channel, we derived the OPE of ${O}_-(z_1)\ {O}_-(z_2)$ in \eqref{ope2}. Then we studied the OPE limit of 3 pt function in $13 \leftrightarrows 2$ and $23 \leftrightarrows 1$ channels to show that these channels are also consistent with \eqref{ope2}. We discussed the possible relation of this singular term with the leading singular term in OPE of light transformed gluons. The corresponding light transformed OPE in various channels has been derived in \eqref{lope2}, \eqref{Lope13to2} and \eqref{Lope23to1} respectively. It is perplexing that the OPE coefficients turn out to be channel dependent.

\subsection{Permutation symmetry of the three-point function} \label{Sec 4.3}
If we look at the momentum space three-point correlator, then we find that under a change of 1 and 2 the correlator picks up a negative sign. This problem is carried over in the light transformed amplitude as well \eqref{3L}. This is not expected from a correlator of bosonic particles. The reason we pick up a negative sign is because we are working with the 'colour stripped' amplitude instead of working with the total MHV amplitude \cite{Taylor:2017sph}. The total momentum space MHV amplitude for the $12\leftrightarrows3$ channel is given by,
\begin{equation}
	\mathcal{M}^{\alpha_{1}\alpha_{2}\alpha_{3}}_{--+}=2^{\frac{3}{2}}\Big(Tr(T^{\alpha_{1}}T^{\alpha_{2}}T^{\alpha_{3}})A_{--+}+Tr(T^{\alpha_{1}}T^{\alpha_{3}}T^{\alpha_{2}})A_{-+-} \Big) 
\end{equation}
where $T^{\alpha_{i}}$ are the generators of $U(n)$ transformations and $\alpha_{i}$ are the group indices or colour indices. The colour stripped amplitudes are functions of $\langle ij\rangle=z_{ij}=z_{i}-z_{j}$. We already have an expression for $A_{--+}$ with colour order (123 )and we need to determine $A_{-+-}$ which has colour order (132). This can be easily obtained by using the general expression for the colour stripped amplitude \cite{Taylor:2017sph},
\begin{equation}
	A_{s_{1}s_{2}s_{3}}= \langle 12\rangle ^{d_{3}} \langle 23\rangle ^{d_{1}} \langle 31 \rangle^{d_{2}} \label{37}
\end{equation}
where $s_{i}$ are the helicities and 
$$
d_{1}=s_{1}-s_{2}-s_{3} ,\quad d_{2}=s_{2}-s_{1}-s_{3} ,\quad d_{3}=s_{3}-s_{1}-s_{2}.
$$
Using \eqref{37} one can show that $A_{-+-}= -A_{-+-} $ by exploiting the fact that $\langle ij \rangle=-\langle ji\rangle$. Now, following the section 4.1 and 4.2, we find that the Mellin transformed total MHV amplitude of the $12\leftrightarrows3$ is, 
\begin{align}
	\tilde{\mathcal{M}}_{--+}^{\alpha_{1}\alpha_{2}\alpha_{3}}|_{12\leftrightarrows3}=2^{\frac{3}{2}}\pi\delta(\sum_j\lambda_j)\  z_{23}^{i \lambda_1 -1} &z_{13}^{i \lambda_2-1}z_{12}^{i \lambda_3 + 1}\ \delta(\bar{z}_{12})\delta(\bar{z}_{13})\nn\\
	&\prod_j\T(\sigma_{*j})\left(Tr(T^{\alpha_{1}}T^{\alpha_{2}}T^{\alpha_{3}})-Tr(T^{\alpha_{1}}T^{\alpha_{3}}T^{\alpha_{2}})\right) .
\end{align} 
The light transformed amplitude would be,
\begin{align}
	L[\tilde{\mathcal{M}}_{--+}^{\alpha_{1}\alpha_{2}\alpha_{3}}]|_{12\leftrightarrows3}= -2^{\frac{3}{2}}\pi\ sgn(z_{12})\ \delta(\sum_j\lambda_j)\  {z_{32}^{i \lambda_3+i\lambda_1}}\ & \z_{13}^{i \lambda_1}\ \z_{23}^{i \lambda_2}\ z_{31}^{-i \lambda_1}\ \f{\Gamma(i\lambda_1)\ \Gamma(i\lambda_2)}{\Gamma(-i\lambda_3)}\nn\\
	&\left(Tr(T^{\alpha_{1}}T^{\alpha_{2}}T^{\alpha_{3}})-Tr(T^{\alpha_{1}}T^{\alpha_{3}}T^{\alpha_{2}})\right) .
\end{align}
Where it is understood that $L[\tilde{\mathcal{M}}_{--+}]$ implies taking the Light transform of the individual colour stripped amplitudes. One can check that this total light transformed celestial amplitude is symmetric under the exchange of 1 and 2. 
We also find that under 1 and 2 exchange the two channels, $13\leftrightarrows2$ and $23\leftrightarrows1$ , are carried into each other so if we sum over all channels, then the total three-point function remains invariant under permutation symmetry.

\section{Four-Point Function of Gluons } \label{Sec5}
In section \ref{Sec 4.3} we saw that the light transformed three-point function was not invariant under $1 \rightleftarrows 2$ which led us to use the total MHV amplitude. Following the analysis of section 4.2 we begin our discussion of four-point function by considering the total MHV amplitude \cite{Parke:1986gb},\cite{Stieberger:2018onx} which can be written as a sum of two colour stripped amplitudes $A_{--++}$ with colour order (1234) and $A_{-+-+}$ with colour order (1324)\cite{DelDuca:1999rs}, 
\begin{align}
	\mathcal{M}^{\alpha_{1}\alpha_{2}\alpha_{3}\alpha_{4}}_{--++}=  8\Big[&\Big(Tr(1234)+Tr(1432) -
	Tr(1243)-Tr(1342)\Big)A_{--++}+
	\Big(Tr(1324)+Tr(1423)- \nn\\
	&Tr(1342)-
	Tr(1243)\Big)A_{-+-+}\Big]. \label{MHV}
\end{align} 
where we used $Tr(T^{\alpha_{i}}T^{\alpha_{j}}T^{\alpha_{k}}T^{\alpha_{l}})\equiv Tr(ijkl)$ for compactification of the expressions. In terms of the celestial coordinates and energy variables,
\begin{eqnarray}
A_{--++} = \delta(\sum_{j=1}^4\lambda_j) \Big(\f{E_1E_2}{E_4E_3} \f{z_{12}^3}{z_{23}z_{34}z_{41}} \Big), ~~~ A_{-+-+} =  \delta(\sum_{j=1}^4\lambda_j) \Big(\f{E_1E_2}{E_4E_3} \f{z_{12}^4}{z_{13} z_{32}z_{24}z_{41}} \Big).
\end{eqnarray}
The Mellin transforms of the colour ordered $A_{--++}$ \cite{Pasterski:2017ylz} is given by,  
\begin{equation}
	\tilde A_{--++}=(-\epsilon_1\epsilon_4)^{i\lambda_1}(\epsilon_2\epsilon_4)^{i\lambda_2}(-\epsilon_3\epsilon_4)^{i\lambda_3}\Big[-\f{\pi}{2}\delta(\sum_j\lambda_j)\ \delta(|z-\bar{z}|) \big[\prod_{i < j}^4
	z_{ij}^{\f{h}{3}-h_i-h_j}\z_{ij}^{\f{\bar{h}}{3}-\bar{h}_i-\bar{h}_j}\big]\ z^{5/3}(1-z)^{-1/3}\prod_{j=1}^4\T(\sigma_{*j})\Big], \label{64}
\end{equation}
where $\epsilon_i=1 (-1)$ for outgoing (incoming) particles. 
However our experience with dealing with three-point functions tell us that we should consider a sum over all channels of the amplitude, this give us \footnote{We recall our phase convention $(-1)^{-i\lambda}=(-1)^{i\lambda}$, as used in \cite{Pasterski:2017ylz} and \cite{Fan:2021isc}.}
\begin{align}
	\tilde A_{--++}=-\f{\pi}{2}\delta(\sum_{j=1}^4\lambda_j)\ \delta(|z-\bar{z}|) \big[\prod_{i < j}^4
	z_{ij}^{\f{h}{3}-h_i-h_j}\z_{ij}^{\f{\bar{h}}{3}-\bar{h}_i-\bar{h}_j}\big]\ z^{5/3}(1-z)^{-1/3} \label{74}.
\end{align}
Similarly one finds that,
\begin{align}
	\tilde A_{-+-+}= \frac{\pi}{2}\delta(\sum_{j=1}^4\lambda_j)\ \delta(|z-\bar{z}|)\ [\prod_{i < j}^4
	z_{ij}^{\f{h}{3}-h_i-h_j}\z_{ij}^{\f{\bar{h}}{3}-\bar{h}_i-\bar{h}_j}]\ z^{8/3}(1-z)^{-1/3} \label{300}
\end{align}
We note that $\tilde A_{-+-+}=-z\tilde A_{--++}$. As a consequence of this, the total colour ordered amplitude for this channel \eqref{MHV} is covariant under the permutation $1 \leftrightarrow 2$:
\begin{align}
	\mathcal{M}^{\alpha_{1}\alpha_{2}\alpha_{3}\alpha_{4}}_{--++} = - \mathcal{M}^{\alpha_{2}\alpha_{1}\alpha_{3}\alpha_{4}}_{-+-+}  \label{MellnColour}
\end{align}

\subsection{Light Transformed Four-Point Function}
%

It is possible to take the light transform in various combinations \cite{Hu:2022syq} but one of the main motivations is to make the four-point function non-distributional. One may ideally choose to light transform all four operators. However as the first step, for the purpose of getting rid of ultra-locality (delta function dependence), one may choose to work with at least one of the operators to be light transformed. We choose to light transform the third operator: 
\begin{align}
	&\langle{O}_{-}(z_1,\z_1)\ O_-(z_2,\z_2)\ L^+[O_+](z_3,\z_3)\ O_+(z_4,\z_4)\rangle\ \ =\ \int dz'_3 \f{ \ \tilde A_{--++}(z_1,z_2,z_3',z_4)}{(z_3-z_3')^{2-2h_3}}  
\end{align}\\
where $\tilde{A}_{--++}(z_1,z_2,z_3',z_4)$ is given in \eqref{74}. We solve the integral using the delta function prescription given in Appendix \ref{AppA1}. We use the constraint given by $z=\z$ which leads to,
$$z^{*}_3=\frac{z_1  \z\ z_{24}+z_4 z_{12}}{ \z\ z_{24}+z_{12}}.$$
We write our result in terms of the new conformal dimensions which are given by $$ h_1= \frac{i\lambda_1}{2}, h_2= \frac{i\lambda_2}{2}, h_3= -\frac{i\lambda_3}{2}, h_4= 1+\frac{i\lambda_4}{2}$$
\be \bar{h}_1= 1+\frac{i\lambda_1}{2}, \bar{h}_2= 1+\frac{i\lambda_2}{2}, \bar{h}_3= \frac{i\lambda_3}{2}, h_4= \frac{i\lambda_4}{2} \label{dim} \ee 
Hence the light transform of the four point correlator is given by, 
\begin{align}
	&\langle{O}_{-}(z_1,\z_1)\ O_-(z_2,\z_2)\ L^+[O_+](z_3,\z_3)\ O_+(z_4,\z_4)\rangle\ 
	=-\f{\pi}{2}\delta(\sum_j\lambda_j)\ [\prod_{\substack{{i < j}}}^4z_{ij}^{\f{h}{3}-h_i-h_j}\z_{ij}^{\f{\bar{h}}{3}-\bar{h}_i-\bar{h}_j}
	]\   \times \nn\\
	&z^{2/3+\f{i\lambda_4}{2}-\f{i\lambda_3}{6}}\  \z^{1/3+\f{i\lambda_2}{2}+\f{i\lambda_1}{2}}\ (1-\z)^{-2/3-\f{i\lambda_2}{2}-\f{i\lambda_3}{2}}\ (1-z)^{-1/3-\f{i\lambda_2}{2}-\f{i\lambda_3}{6}} \left(z-\z\right)^{i \lambda_3} \label{4ptL}
\end{align}
Note that \eqref{4ptL} is no more ultra-local on the celestial sphere. \color{black}  Following the same analysis we find that the light transform of $\tilde{A}_{-+-+}$ \eqref{300},
\begin{align}
	&\langle{O}_{-}(z_1,\z_1)\ L^+[O_+](z_3,\z_3)\ O_-(z_2,\z_2)\ O_+(z_4,\z_4)\rangle\ 
	=\f{\pi}{2}\delta(\sum_j\lambda_j)\ [\prod_{\substack{{i < j}}}^4z_{ij}^{\f{h}{3}-h_i-h_j}\z_{ij}^{\f{\bar{h}}{3}-\bar{h}_i-\bar{h}_j}
	]\   \times \nn\\
	&z^{2/3+\f{i\lambda_4}{2}-\f{i\lambda_3}{6}}\  \z^{4/3+\f{i\lambda_2}{2}+\f{i\lambda_1}{2}}\ (1-\z)^{-2/3-\f{i\lambda_2}{2}-\f{i\lambda_3}{2}}\ (1-z)^{-1/3-\f{i\lambda_2}{2}-\f{i\lambda_3}{6}} \left(z-\z\right)^{i \lambda_3} \label{4ptL1}
\end{align}

Thus the four point function resembles four point functions in ordinary CFT's. In the next subsection, we discuss the permutation symmetry of four-point function.

 \subsection{Permutation symmetry}
The total momentum space MHV four-point amplitudes are invariant under all particle exchanges, individually for each channel. We have seen earlier that the total Mellin amplitude remains invariant upto a sign under particle exchange. In the following, we analyse the effect of particle exchange on the light transformed amplitude keeping in mind that the light transformed amplitudes are analytically well defined when all channel contributions are summed over.

We begin with the transformation property of the colour stripped $A_{--++}$ channel summed light transformed amplitude under this exchange,
\begin{align}
	&\
	\langle{O}_{-}(z_1)\ O_-(z_2)\ L^+[O_+](z_3)\ O_+(z_4)\rangle|_{1 \leftrightarrow 2}=-\frac{\pi}{2}\delta(\sum_j\lambda_j)\ [\prod_{\substack{{i < j}}}^4z_{ij}^{\f{h}{3}-h_i-h_j}\z_{ij}^{\f{\bar{h}}{3}-\bar{h}_i-\bar{h}_j}
	] \times \nn\\
	&z^{2/3+\f{i\lambda_4}{2}-\f{i\lambda_3}{6}}\  \z^{1/3+\f{i\lambda_2}{2}+\f{i\lambda_1}{2}}\ (1-\z)^{1/3-\f{i\lambda_2}{2}-\f{i\lambda_3}{2}}\ (1-z)^{-1/3-\f{i\lambda_1}{2}-\f{i\lambda_4}{2}-\f{2i\lambda_3}{3}} \left(z-\z\right)^{i \lambda_3}\label{LT12}.
\end{align}
We find that the light transformed amplitude changes in a non-trivial manner and the difference is not just a phase factor.  One can arrive at a similar concluson for colour stripped $A_{-+-+}$ where one would have, 
\begin{align}
	&\
	\langle{O}_{-}(z_1)\ O_-(z_2)\ L^+[O_+](z_3)\ O_+(z_4)\rangle|_{1 \leftrightarrow 2}=-\frac{\pi}{2}\delta(\sum_j\lambda_j)\ [\prod_{\substack{{i < j}}}^4z_{ij}^{\f{h}{3}-h_i-h_j}\z_{ij}^{\f{\bar{h}}{3}-\bar{h}_i-\bar{h}_j}
	] \times \nn\\
	&z^{2/3+\f{i\lambda_4}{2}-\f{i\lambda_3}{6}}\  \z^{1/3+\f{i\lambda_2}{2}+\f{i\lambda_1}{2}}\ (1-\z)^{1/3-\f{i\lambda_2}{2}-\f{i\lambda_3}{2}}\ (1-z)^{-1/3-\f{i\lambda_1}{2}-\f{i\lambda_4}{2}-\f{2i\lambda_3}{3}} \left(z-\z\right)^{i \lambda_3}\label{Channel2LT12}.
\end{align}
To circumvent this problem we resort to the colour summed amplitude and we find that, 
\begin{align}
	&\mathcal{L}[\mathcal{M}^{\alpha_{1}\alpha_{2}\alpha_{3}\alpha_{4}}_{--++}]|_{1\leftrightarrow2}= 8\pi\delta(\sum_j\lambda_j)\ [\prod_{\substack{{i < j}}}^4z_{ij}^{\f{h}{3}-h_i-h_j}\z_{ij}^{\f{\bar{h}}{3}-\bar{h}_i-\bar{h}_j}
	] \ z^{2/3+\f{i\lambda_4}{2}-\f{i\lambda_3}{6}}
 	\z^{1/3+\f{i\lambda_2}{2}+\f{i\lambda_1}{2}} (1-\z)^{-2/3-\f{i\lambda_2}{2}-\f{i\lambda_3}{2}}\times\nn\\ 
	& (1-z)^{-1/3-\f{i\lambda_2}{2}-\f{i\lambda_3}{6}} \left(z-\z\right)^{i \lambda_3}\big(Tr(1234)+Tr(1432)-Tr(1243-Tr(1342)\big)
	-\bar{z}\big(Tr(1324)+Tr(1423)\nn\\
&	-Tr(1243-Tr(1342)\big)
\end{align}
 This implies  $\mathcal{L}[\tilde{M}^{\alpha_{1}\alpha_{2}\alpha_{3}\alpha_{4}}_{--++}|]_{1\leftrightarrow2}=-\mathcal{L}[\tilde{M}^{\alpha_{2}\alpha_{1}\alpha_{3}\alpha_{4}}_{--++}]$, that is we have this symmetry as one would expect from \eqref{MellnColour}.
\section{Conclusion}
In conclusion, we studied some desirable features of the light transformed operators apart from the fact that it helps us in writing the momentum space amplitudes in a different basis. Light transform represents a change of basis and we find the light transformed correlators are not only covariant under SL(2,C) transformations but are also invariant under translation on the celestial sphere and we have demonstrated this for the two-point and three-point function. Further, we note that light transform of colour stripped amplitudes do not always enjoy 'particle exchange' symmetry. However if we light transform the entire MHV amplitude with colour factors included, we get this desired property. 

We also showed existence of a peculiar channel dependent term in the light transformed three point function. Present in the $13\leftrightarrows2$ and the
$23\leftrightarrows1$ channels, this term breaks special conformal
covariance of these channels individually as seen in \eqref{f} and \eqref{f2}. However, we showed that when
contributions of the channels are summed up, this anomaly
cancels.  The interpretation of such correlators (with sum over all channels) remains to be understood vis-a-vis the state operator correspondence. The origin of the anomalous term via the definition
(prescription) of line integral for the light transform has been discussed in Appendix \ref{AppB}. 

Using the three-point functions we obtained the leading OPE between the gluon operators in \eqref{ope2} and light transformed gluon operators. There seems to be a direct map between the leading singularity in OPE of gluon operators and the leading singularity in the OPE of light transformed gluon operators. But such a direct map does not exist between the respective OPE coefficients. The OPE of light transformed in various channels has been derived in \eqref{lope2}, \eqref{Lope13to2} and \eqref{Lope23to1} respectively. We obtained a rather strange result that the OPE coefficients are channel dependent. This calls for a further scrutiny, particularly associativity of the algebra of light transformed gluon CCFT operators needs to be checked. Notably, a recent work \cite{Ren:2022sws} delved in to the topic of associativity of CCFT operator algebra for the case of gravitons and gluons.

We obtained the four point function involving one light transformed gluon in section \ref{Sec5} ad showed that it has the usual CFT power law behaviour. We also showed that full amplitude with all gluon colour permutations is required to have a particle exchange symmetric correlation function.

Other questions that we aim to tackle in the future are as follows : 
\begin{itemize}
	\item The presence of $sgn(z_{12})$ in the three point-function of the light transformed operators is intriguing. It is related to a similar discontinuity in the OPE's of the celestial operators we studied in \eqref{ope2} and \eqref{Lope13to2}. All such possible OPE structures for CCFT's need to be studied and classified.
	\item It is fair to say that the complete significance of light transformed operators in CCFTs is not clear yet. It remains to be understood if the light transformed operators satisfy state operator correspondence. A related question is to understand if there is a sense in which light transformed operators can be treated as local operators. It needs to be investigated if the corresponding four-point function satisfies cluster decomposition.
	\item Various OPE limits of the light transformed four-point functions should be studied systematically to get a better understanding of the operator algebra of CCFTs in the light transformed basis. These OPE's need to be compared with the OPE's that we derived from the 3 pt function. 
\end{itemize}

\section*{Acknowledgement}
Research of RB is supported by the following grants from the SERB, India: CRG/2020/002035, SRG/2020/001037. SB is also supported by the grant by SERB, India: SRG/2020/001037. 
\appendix

\section{Delta function integral in Euclidean signature} \label{AppA1}
Let us consider following integral that appears in the light transformed two point function :
\be
\oint_{\z_1}\f{d\z_1'}{(\bar{z_1}-\bar{z}'_1)^{2-2\bar{h}_1}}\  \delta(\z_1'-\z_2)
\ee
The delta function can be defined by \cite{Donnay:2020guq} 
\begin{itemize}
	\item For $Re(\z'_1-\z_2)>0$ 
	\be
	\delta(\z_1'-\z_2)=\lim_{\nu\rightarrow 0} \nu^{-(\z_1'-\z_2)}\Gamma(\z_1'-\z_2)
	\ee
	Let us use above expression to do the integral :
	\be
	\lim_{\nu\rightarrow 0}\oint_{\z_1}\f{d\z_1'}{(\bar{z_1}-\bar{z}'_1)^{2-2\bar{h}_1}}\  \nu^{-(\z_1'-\z_2)}\Gamma(\z_1'-\z_2)
	\ee
	We deform the contour and pick out the poles of the Gamma function to get 
	\be
	\lim_{\nu\rightarrow 0}\ \sum_{n=0}^\infty \f{(-1)^n}{n!}\ \f{ \nu^{n}}{(\z_1-\z_2+n)^{2-2\bar{h}_1}}\ \ =\  \f{1}{(\z_{12})^{2-2\bar{h}_1}}\ .
	\ee
	\item For $Re(\z'_1-\z_2)<0$ 
	\be
	\delta(\z_1'-\z_2)=\lim_{\nu\rightarrow 0} \nu^{(\z_1'-\z_2)}\Gamma(\z_2-\z_1')
	\ee
	Let us use above expression to do the integral :
	\be
	\lim_{\nu\rightarrow 0}\oint_{z_1}\f{d\z_1'}{(\bar{z_1}-\bar{z}'_1)^{2-2\bar{h}_1}}\  \nu^{(\z_1'-\z_2)}\Gamma(\z_2-\z_1')
	\ee
	Like in the previous case we will deform the contour and pick out the poles of the Gamma function to get 
	\be
	\lim_{\nu\rightarrow 0}\ \sum_{n=0}^\infty \f{(-1)^n}{n!}\ \f{ \nu^{n}}{(\z_1-\z_2+n)^{2-2\bar{h}_1}}\ \ =\  \f{1}{(\z_{12})^{2-2\bar{h}_1}}\ .
	\ee
\end{itemize}

\section{Conformal transformation} \label{AppB}
Let us discuss the conformal transformation of various channels of the 3pt scattering process. We start with the integral for the $12\rightarrow 3$ channel given in \eqref{123} :
\begin{align}
I_{12\rightarrow 3}(z_1,z_2,z_3)= \int_{z_2}^{z_1} d x \f{sgn(z_{12})}{(z_3-x)^{-i \lambda_3}}\ (z_{2}-x)^{-1+i \lambda_1} (z_{1}-x)^{-1+i \lambda_2}z_{12}^{1+i \lambda_3}\ .
\end{align}
Here we will focus on the transformation of $z$ variable keeping $\z$ fixed so we have not included the barred terms in above expression. We have for $z_{i}'=\f{az_i +b}{cz_i +d}$ : 
\begin{align}
I_{12\rightarrow 3}(z'_1,z'_2,z'_3)= \int_{z'_2}^{z'_1} d x \f{sgn(z'_{12})}{(z'_3-x)^{-i \lambda_3}}\ (z'_{2}-x)^{-1+i \lambda_1} (z'_{1}-x)^{-1+i \lambda_2}z_{12}'^{1+i \lambda_3}\\ .
\end{align}
Next let us change the variable to $x=\f{ax' +b}{cx' +d}$ to get 
\begin{align}
&I_{12\rightarrow 3}(z'_1,z'_2,z'_3)= \int_{z_2}^{z_1} \f{d x'}{(a-cx')^2} \f{sgn(z'_{12})}{(z'_3-x')^{-i \lambda_3}}\ (z'_{2}-\f{ax' +b}{cx' +d})^{-1+i \lambda_1} (z'_{1}-\f{ax' +b}{cx' +d})^{-1+i \lambda_2}z_{12}'^{1+i \lambda_3}\nn .
\end{align}
 Using $z_{ij}'=\f{z_{ij}}{(cz_i +d)(cz_j +d)}$, we get
\begin{align}
&I_{12\rightarrow 3}(z'_1,z'_2,z'_3)=\nn\\
& \f{sgn(cz_1+d)}{(cz_1+d)^{i\lambda_3+i\lambda_2}}\f{sgn(cz_2+d)}{(cz_2+d)^{i\lambda_3+i\lambda_1}}\f{1}{(cz_3+d)^{i\lambda_3}}\int_{z_2}^{z_1}  \f{d x'}{(cx'+d)^2} \f{sgn(z_{12})}{(z_3-x')^{-i \lambda_3}}\ \f{(z_{2}-x')^{-1+i \lambda_1}}{(cx'+d)^{-2}} (z_{1}-x')^{-1+i \lambda_2}z_{12}^{1+i \lambda_3}\nn .\\
&= \f{sgn(cz_1+d)}{(cz_1+d)^{i\lambda_3+i\lambda_2}}\f{sgn(cz_2+d)}{(cz_2+d)^{i\lambda_3+i\lambda_1}}\f{1}{(cz_3+d)^{i\lambda_3}}\int_{z_2}^{z_1}  dx' \f{sgn(z_{12})}{(z_3-x')^{-i \lambda_3}}\ (z_{2}-x')^{-1+i \lambda_1} (z_{1}-x')^{-1+i \lambda_2}z_{12}^{1+i \lambda_3}
\end{align}
Hence we see that above integral has the right transformation under SL(2,R) transformations of $z_i$ :
\begin{align}
&I_{12\rightarrow 3}(z'_1,z'_2,z'_3)= sgn(cz_2+d)\ sgn(cz_1+d)\ \prod_{i=1}^3\f{1}{(cz_i+d)^{2h_i}} \ I_{12\rightarrow 3}(z_1,z_2,z_3)
\end{align}
This is consistent with the transformation of the 3 pt correlator where $h_i$ refer to the conformal weights of the light tranformed operators in the 3pt function in \eqref{3L}. $sgn(cz_2+d)\ sgn(cz_1+d)$ appears due to the presence of the $sgn(z_{12})$ function. 

Next we consider the integral in \eqref{132} for the $13\rightarrow 2$ channel :
\begin{align}
I_{13\rightarrow 2}(z_1,z_2,z_3)=  z_{12}^{1+i \lambda_3}\int_{z_2}^{\infty}  \f{d x}{(z_3-x)^{-i \lambda_3}}\ (z_{2}-x)^{-1+i \lambda_1} (z_{1}-x)^{-1+i \lambda_2} .\label{99}
\end{align}
We have not written down the $sgn$ function as it is not relevant for the subsequent discussion. Following the steps as before, we get 
\begin{align}
I_{13\rightarrow 2}(z_1',z_2',z_3')= \prod_{i=1}^3\f{1}{(cz_i+d)^{2h_i}} \ z_{12}^{1+i \lambda_3}\int_{z_2}^{-\f{d}{c}}  \f{d x'}{(z_3-x')^{-i \lambda_3}}\ (z_{2}-x')^{-1+i \lambda_1} (z_{1}-x')^{-1+i \lambda_2}.\label{98}
\end{align}
We cannot express above integral in terms of the original integral in \eqref{99} due to the presence of the upper limit. Hence the integral is not covariant. This is the reason why the second term in the three-point function corresponding to the $13\leftrightarrows 2$ channel in \eqref{L132} is not covariant. We find that the three-point function satisfies the Ward identities for translation and scaling. In \eqref{98}, this is clear when we note that translation and scaling have $c=0$ hence the integral is also covariant. The special conformal transformations have a non zero $c$ and hence the corresponding Ward identity is not satisfied as seen in \eqref{sct}.
\bibliographystyle{unsrt}
\section{Special Conformal Transformation of Three Point Function} \label{AppC}
We have seen in Section \ref{Sec4.4} that the Light transformed three point correlator does not satisfy the ward identity for the Special Conformal Transformation and we have justified this from a general point of view in Appendix \ref{AppB}. In this section we provide the computation details which led to \eqref{sct}, we start by using the following identity for the hypergeometric function of \eqref{3L132}
\begin{equation}
	{}_2F_1(1,i(\lambda_1+\lambda_2),1+i\lambda_2,w)=(1-w)^{-i\lambda_1}{}_2F_1(1-i\lambda_1,i\lambda_2,1+i\lambda_2,w) \quad \text{with} \quad w=\frac{z_{32}}{z_{12}}\label{102}
\end{equation}
and we have use the fact that ${}_2F_1(a,b,c,w)={}_2F_1(b,a,c,w)$. We note that the right side of \eqref{102} is of the form $(1-z)^{a-1}{}_2F_1(a,b,b+1,z)$. So we can use the definition of the hypergeometric function as a Gauss series to simplify this function as follows,
\begin{align}
{}_2F_1(a,b,b+1,w)&= \frac{b}{\Gamma(a)}\sum_{s}\frac{\Gamma(a+s)}{(b+s)s!}w^s \nn\\
& = bw^{-b}\int dw \Big(\sum_{s}\frac{\Gamma(a+s)}{\Gamma(a)s!} z^{s} \Big) w^{b-1}\nn\\
&= bw^{-b}\int dw\ (1-w)^{-a}w^{b-1} \label{103}.
\end{align}
Now, let us define, 
\begin{equation}
	g= (1-w)^{a-1}bw^{-b}\int dw \ (1-w)^{-a}w^{b-1} \quad \text{and} \quad G=w_{12}^{i\lambda_3}h. \label{104}
\end{equation}
To evaluate the left side of equation \eqref{sct} we first calculate the derivatives of g with respect to the $z_{i}$'s and they are, 
\begin{align}
	& \partial_{1}G= z_{12}^{i\lambda_3-1}\big[i\lambda_3 -(z\partial_{z})\big]g \nn\\
	& \partial_{2}G= -z_{12}^{i\lambda_3-1}\big[i\lambda_3+(1-z)\partial_{z}\big]g\nn\\
	& \partial_{3}G= z_{12}^{i\lambda_3-1} \partial_{z}g \label{105}
\end{align}
We evaluate $ \sum_{i}(z_{i}^2\partial_{i}G + 2h_{i}z_{i}G)$ using the above equations and after some algebraic simplification it turns out to be, 
\begin{equation}
\sum_{i}z_{i}^2\partial_{i}G + 2h_{i}z_{i}G=-i\lambda_2(z_1-z_2)^{i\lambda_3+1}.
\end{equation}
where we note that $h_{i}$ in terms of $\lambda_{i}$ are given by \eqref{hl3}.

\bibliography{cel}

\end{document}